\documentclass[twocolumn,showpacs,noshowkeys,preprintnumbers,amsmath,amssymb,superscriptaddress,prb,floatfix]{revtex4}
\usepackage{graphicx,tabularx}
\usepackage{amssymb}
\usepackage[mathcal]{euscript}
\newcommand{\abs}[1]{\left\vert#1\right\vert}

\newcommand{\lit}{\textrm{L\lowercase{i}A\lowercase{l}H}_4}
\newcommand{\m}{\textrm{MA\lowercase{l}H}_4}
\newcommand{\md}{\textrm{M}'\textrm{(A\lowercase{l}H}_4\textrm{)}_2}
\newcommand{\mm}{\textrm{M}_3\textrm{A\lowercase{l}H}_6}
\newcommand{\litt}{\textrm{L\lowercase{i}}_3\textrm{A\lowercase{l}H}_6}
\newcommand{\naa}{\textrm{N\lowercase{a}}_3\textrm{A\lowercase{l}H}_6}
\newcommand{\na}{\textrm{N\lowercase{a}A\lowercase{l}H}_4}
\newcommand{\mg}{\textrm{M\lowercase{g}(A\lowercase{l}H}_4\textrm{)}_2}
\newcommand{\Am}{\AA $\;$}
\newcommand{\nah}{\textrm{N\lowercase{a}H}}
\newcommand{\lih}{\textrm{L\lowercase{i}H}}
\newcommand{\mgh}{\textrm{M\lowercase{g}H}_2}
\newcommand{\alh}{\textrm{A\lowercase{l}H}_3}
\newcommand{\alcl}{\left(\textrm{A\lowercase{l}H}_4\right)^-}
\newcommand{\alcll}{\left(\textrm{A\lowercase{l}H}_6\right)^{3-}}
\newcommand{\q}{\mathbf{q}}
\newcommand{\limq}{\lim_{|\q|\to0}}
\newcommand{\bk}{\mathbf{k}}
\newcommand{\bracet}[2]{\left<#1\vert#2\right>}

\begin{document}
\title{Electronic structure and optical properties of lightweight metal hydrides}
\author{M. J. van Setten}
\affiliation{Electronic Structure of Materials, Institute for Molecules and Materials, Faculty
of Science,\\ Radboud University Nijmegen, Toernooiveld 1, 6525 ED Nijmegen, The Netherlands}
\author{V. A. Popa}
\affiliation{Computational Materials Science, Faculty of Science and Technology and MESA+
Research Institute, University of Twente, P.O. Box 217, 7500 AE Enschede, The Netherlands}
\author{G. A. de Wijs}
\affiliation{Electronic Structure of Materials, Institute for Molecules and Materials, Faculty
of Science,\\ Radboud University Nijmegen, Toernooiveld 1, 6525 ED Nijmegen, The Netherlands}
\author{G. Brocks}
\affiliation{Computational Materials Science, Faculty of Science and Technology and MESA+
Research Institute, University of Twente, P.O. Box 217, 7500 AE Enschede, The Netherlands}

\date{\today}



\begin{abstract}
We study the electronic structures and dielectric functions of the simple hydrides $\lih$, $\nah$,
$\mgh$ and $\alh$, and the complex hydrides $\litt$, $\naa$, $\lit$, $\na$ and $\mg$, using first
principles density functional theory and $GW$ calculations. All these compounds are large gap
insulators with $GW$ single particle band gaps varying from 3.5 eV in AlH$_3$ to 6.5 eV in the
MAlH$_4$ compounds. The valence bands are dominated by the hydrogen atoms, whereas the conduction
bands have mixed contributions from the hydrogens and the metal cations. The electronic structure
of the aluminium compounds is determined mainly by aluminium hydride complexes and their mutual
interactions. Despite considerable differences between the band structures and the band gaps of the
various compounds, their optical responses are qualitatively similar. In most of the spectra the
optical absorption rises sharply above 6 eV and has a strong peak around 8 eV. The quantitative
differences in the optical spectra are interpreted in terms of the structure and the electronic
structure of the compounds.
\end{abstract}

\maketitle

\section{Introduction}

The large scale utilization of hydrogen as a fuel crucially depends on the development of compact
hydrogen storage materials with a high mass content of hydrogen.\cite{bog_rev} Hydrides of group
I-III metals in the upper rows of the periodic table could meet this requirement. These metals are
sufficiently light for their hydrides to have a large gravimetric hydrogen density; for instance,
$\mgh$ contains 7.7 wt.\% hydrogen. One must be able to extract hydrogen at a moderate temperature,
however, and therefore the metal hydride should be neither too stable nor too unstable. Simple
metal hydrides do not satisfy this demand. For example, the binding energy of $\mgh$ is too
large,\cite{griess,bohm} whereas the binding energy of $\alh$ is close to zero.\cite{wolverton}

This has stimulated research into binary intermetallic hydrides such as the alanates $\m$, $\md$,
with M and M$'$ a light alkali and alkaline earth metal, respectively. Some of the properties of
these compounds have indeed improved as compared to the simple hydrides, but the compound that
meets both the stability and the storage capacity demands has not been found yet. Whereas sodium
alanate, $\na$, releases hydrogen in two reaction stages with enthalpies close to the ideal value,
its active gravimetric hydrogen density is only 5.5 wt.\%.\cite{bog_rev,bog,bog2} Magnesium
alanate, $\mg$, and lithium alanate, $\lit$, have a higher active gravimetric hydrogen density of
7.0 and 8.0 wt.\%, respectively. However, they are not sufficiently stable with respect to
decomposition into simpler hydrides.\cite{vansetten,lov,lovvik1,andreasen05,mamatha06,dymo94}

A suitable ternary intermetallic hydride might satisfy all requirements. The number of possible
ternary compounds is very large, however, and searching for the optimal composition becomes very
tedious, unless one uses a combinatorial approach. Such a technique has been proposed recently, in
which thin films are grown with tunable composition gradients.\cite{dam} The composition is then a
function of the position on the film. This avoids having to synthesize all compositions
individually, but one still needs to be able to identify the most promising ones. It is proposed
that identification can be based upon the optical properties of suitable metal hydrides being very
different from those of their host metals.\cite{dam} This has first been demonstrated conclusively
for YH$_3$,\cite{huiberts96} and since then for a number of other so-called ``switchable mirror"
rare earth and transition metal
compounds.\cite{kremers98,kerssemakers00,richardson01,lohstroh04,lohstroh04-2,kumar05,lokhorst05}
The compounds that absorb the maximum amount of hydrogen, become semiconductors or insulators.

If one applies this technique to group I-III metal hydrides, it is relevant to know how the optical
properties of these compounds depend on their composition and structure. In this paper we report a
systematic first principles study of the band gaps, the electronic structure and the optical
properties of group I-III metal hydrides. Band gaps and single particle excitations are calculated
within the $GW$ quasi-particle approach; optical excitations are obtained using the random phase
approximation (RPA). We focus upon a number of elements that are of interest for lightweight metal
hydrides, i.e., Li, Na, Mg and Al. In particular, we consider the series of simple metal hydrides
LiH, NaH, $\mgh$, $\alh$ and the binary metal hydrides $\litt$, $\naa$, $\lit$, $\na$ and
$\mg$.\cite{bog2,lov,dilt,dymo,gros,zut,fich,arr,chou} The trends in the optical spectra and
electronic structure are discussed and interpreted in terms of the structure and bonding of the
materials.

This paper is organized as follows. In Sec.~II we outline the computational methods used in our
study. The results are presented in Sec.~III, first for the simple hydrides, then for the binary
$\mm$ hydrides and finally for the $\m$ alanates. Secs.~IV and V contain the discussion and a
summary.

\section{Computational methods}

The results discussed in this paper are obtained using a combination of density functional theory
(DFT) and $GW$ calculations. DFT is used at GGA level to optimize the ground state structure and 
obtain single particle wave functions to be used in the calculation of the optical response. $GW$
is used to generate single particle excitation energies within the quasi-particle (QP) approximation,
 starting from DFT/LDA wavefunctions and eigenvalues.
The optical response is given by the frequency dependent
dielectric function, which is calculated within the random phase approximation (RPA). In the latter
we use single particle wave functions and excitation energies and neglect exciton and local field
effects. The main difference between the DFT and the QP excitation spectra is the size of the
fundamental band gap between occupied and unoccupied states, whereas the dispersion of the bands is
quite similar. We use a scissors operator to the DFT eigenvalues to approximate QP excitation
energies on a dense grid in the Brillouin zone, which is required to calculate the dielectric
function.

\subsection{DFT calculations}

DFT total energies are calculated with the PW91 generalized gradient approximation (GGA)
functional\cite{gga} and the projector augmented wave (PAW) method,\cite{paw,blo} as
implemented in the Vienna \em Ab initio \em Simulation Package
(VASP).\cite{vasp1,vasp2,vasp3} We use standard frozen core PAW potentials and a plane wave
basis set with a kinetic energy cutoff of $312$~eV. The tetrahedron scheme is applied for the
Brillouin zone integration using $\mathbf{k}$-point meshes with a spacing between 0.01 and
0.03 \AA$^{-1}$. The cell parameters and the atomic positions within a unit cell are
optimized by minimizing the total energy, except for the alkali alanates, where only the
atomic positions are optimized and the cell parameters are taken from experiment. These
optimized structures are used as input for the $GW$ calculations.

\subsection{$GW$ calculations}

DFT calculations generally give good results for ground state properties, but not for excited
states. The electronic band gap, for instance, can be underestimated by $\sim 50$\%, and even more
than that for small band gap materials. This stems from an unjustified interpretation of the DFT
(Kohn-Sham) eigenvalues as single particle excitation energies. The latter can be obtained from the
quasi-particle (QP) equation, which involves the non-local, energy dependent self-energy $\Sigma$,
\begin{eqnarray}
\left[-\frac{1}{2}\nabla^{2}+v_{\mathrm{ext}}(\mathbf{r})+V_{H}(\mathbf{r})\right]
                                                       \psi_{n\mathbf{k}}(\mathbf{r})
&+&   \nonumber \\
  + \int d\mathbf{r}'\Sigma(\mathbf{r},\mathbf{r}';\epsilon_{n\mathbf{k}})
                                                       \psi_{n\mathbf{k}}(\mathbf{r}')
&=&                          \epsilon_{n\mathbf{k}} \psi_{n\mathbf{k}}(\mathbf{r}),
\label{QP_eq}
\end{eqnarray}
where $v_{\mathrm{ext}}$ stands for the sum of all nuclear or ionic potentials, $V_{H}$ is the
electrostatic or Hartree potential resulting from the electrons, $\psi_{n\mathbf{k}}(\mathbf{r})$
is the QP wave function, and $\epsilon_{n\mathbf{k}}$ is the QP energy, i.e., the single particle
excitation energy. In practice Eq.~(\ref{QP_eq}) is solved using a number of approximations. The
$GW$ technique approximates the self-energy $\Sigma$ by a dynamically screened exchange
interaction. A large variety of $GW$ implementations exist, in which quite different levels of
approximation are used. We will explain our procedure here and benchmark it on simple hydrides in
the next section.

The $G_0W_0$ approximation is defined by constructing the self-energy $\Sigma$ from the orbitals
and eigenvalues obtained in a DFT calculation within the local density approximation (LDA). If the
QP and LDA wave functions do not differ significantly, i.e.,
$\psi_{n\mathbf{k}}(\mathbf{r})\approxeq\psi_{n\mathbf{k}}^{\mathrm{LDA}}(\mathbf{r})$, then
Eq.~(\ref{QP_eq}) is approximated by\cite{Hybertsen:prb86,Godby:prb88}
\begin{equation}
h_{n\mathbf{k}}+\Sigma_{n\mathbf{k}}(\epsilon_{n\mathbf{k}})=\epsilon_{n\mathbf{k}},
\label{dec_eq}
\end{equation}
where $h_{n\mathbf{k}}=\langle
\psi_{n\mathbf{k}}^{\mathrm{LDA}}|-\frac{1}{2}\nabla^{2}+v_{\mathrm{ext}}+V_{H}|
\psi_{n\mathbf{k}}^{\mathrm{LDA}}\rangle$ and $\Sigma_{n\mathbf{k}}=\langle
\psi_{n\mathbf{k}}^{\mathrm{LDA}}|\Sigma|\psi_{n\mathbf{k}}^{\mathrm{LDA}}\rangle$.
Eq.~(\ref{dec_eq}) is non-linear in $\epsilon_{n\mathbf{k}}$ and it is solved by a root-searching
technique.\cite{QPremark} $GW$ calculations are computationally demanding, so pseudopotentials are
used to represent the ion cores and only the valence electrons are treated explicitly. Calculations
within this scheme have been applied to a wide range of semiconductors and
insulators.\cite{aryasetiawan} They lead to band gaps that are usually within 10\% of the
experimental values, although occasionally somewhat larger deviations are found.\cite{rubio}

In principle, an overlap between core and valence charge densities contributes to the screening,
and thus to the self-energy. This contribution is neglected in a pseudopotential approach, but a
simple estimate of its effect is made by adding
\begin{equation}
(V_{xc}[\rho_v+\rho_c]-V_{xc}[\rho_v])_{n\mathbf{k}}, \label{vxc_eq}
\end{equation}
to the QP energies, where $V_{xc}$ is the LDA exchange-correlation potential, $\rho_{c,v}$ are the
core and valence charge densities, and ${n\mathbf{k}}$ indicates the expectation value with respect
to an LDA wave function as in Eq.~(\ref{dec_eq}).\cite{gelderen1,gelderen2}

The QP equation, Eq.~(\ref{QP_eq}), is not related to DFT, but the scheme outlined above depends on
LDA eigenvalues and wave functions through the $G_0W_0$ approximation for the self-energy and
through the approximation represented by Eq.~(\ref{dec_eq}). The dependence on LDA eigenvalues can
be avoided by constructing the self-energy ($GW$) from QP energies and solve the QP equation
self-consistently. We have previously observed that self-consistency on the eigenvalues is in fact
vital to obtain good results for small band gap semiconductors that are incorrectly described by
LDA as being metallic.\cite{gelderen1,gelderen2} For large band gap materials, however, this
self-consistency does not improve upon the $G_0W_0$ results.\cite{popa} The dependence on LDA wave
functions can be relaxed by solving Eq.~(\ref{QP_eq}) instead of Eq.~(\ref{dec_eq}). However, for
large band gap materials this changes the results only marginally.\cite{popa} The dependence of the
self-energy on the LDA wave functions is not that easily avoided. Self-consistency applied to QP
wave functions worsens the results as compared to the $G_0W_0$ approximation.\cite{popa}

Since all hydrides considered in this paper turn out to be large band gap insulators, we use the
$G_0W_0$ approximation to calculate the QP spectrum. Starting from the optimized structures we
generate wave functions and eigenvalues from an LDA calculation with norm-conserving
pseudopotentials and a plane wave kinetic energy cutoff of 748~eV.\cite{TM} The self-energy
$G_0W_0$ is calculated using the real space, imaginary time
formalism.\cite{gelderen2,rojas,rieger,vanderhorst} We include 350 LDA states, use a real space
grid mesh with a typical spacing of 0.3-0.4 \Am and an interaction cell parameter of $\sim 25$ \Am.
The QP equation is solved in the approximation represented by Eq.~(\ref{dec_eq}). We estimate that
with these parameters QP band gaps are converged numerically to within $\pm 0.02$~eV.

\subsection{Macroscopic dielectric function}

Optical excitations are two-particle excitations, but neglecting excitonic effects they can be
approximated by transitions between single particle states. There is no experimental indication of
strong excitonic effects in metal hydrides. For instance, in LiH the binding energy of the lowest
lying exciton is less than 0.05~eV.\cite{plek,baroni} In this paper we assume that such excitonic
effects can be neglected. In addition we neglect local field effects.

If we consider quasi-particles as independent particles, then the imaginary part of the macroscopic
dielectric function obtains the simple form\cite{adler,wiser,delsole}
\begin{eqnarray}\label{epsimag}
\varepsilon^{(2)}(\hat{\q},\omega)&=&\frac{8\pi^2 e^2}{V} \;
\limq \frac{1}{\abs{\q}^2}\sum_{\bk v c}  \nonumber \\
&& \times \abs{\bracet{u_{c\bk+\q}}{u_{v\bk}}}^2\;
\delta(\epsilon_{c\bk+\q}-\epsilon_{v\bk}-\hbar\omega),
\end{eqnarray}
where $\hat{\q}$ gives a direction, $v\mathbf{k}$ ($c\mathbf{k}$) label single particle states that
are occupied (unoccupied) in the ground state, and $\epsilon$, $u$ are the single particle energies
and the translationally invariant parts of the wave functions; $V$ is the volume of the unit cell.
We have assumed spin degeneracy.

Almost all optical data on hydrides are obtained on micro- or nano-crystalline samples whose
crystallites have a significant spread in orientation. The most relevant quantity then is the
directionally averaged dielectric function, i.e., $\varepsilon^{(2)}(\omega)$ averaged over
$\hat{\q}$. Eq.~(\ref{epsimag}) involves calculating $u_{c\bk+\q}$ for small $\q$ and each $\bk$
and extrapolating to $\q = 0$. Details can be found in Ref.~\onlinecite{kresseps}.

The summation over the Brillouin zone in Eq.~(\ref{epsimag}) is performed using a weighted
tetrahedron scheme.\cite{furt} We found that this scheme allows for a faster convergence with
respect to the number of $\mathbf{k}$-points than various smearing methods. To calculate the
dielectric tensor, we use the same plane wave kinetic energy cutoff and $\mathbf{k}$-point mesh as
for the DFT/GGA calculations. The number of empty bands included is sufficiently large as to
describe all transitions up to at least 16~eV.

If the imaginary part of the dielectric function, $\varepsilon^{(2)}(\omega)$, is calculated for
all frequencies $\omega$, then the real part, $\varepsilon^{(1)}(\omega)$, can be obtained by a
Kramers-Kronig transform. The static component $\varepsilon^{(1)}(0)=\varepsilon^\infty$ can also
be calculated using density functional perturbation theory. Since the latter calculation includes
local field effects, comparing $\varepsilon^\infty$ obtained with the two techniques is a way of
assessing the importance of these effects.\cite{kresseps}

In order to produce the right band gap one should use the $GW$ QP energies in Eq.~(\ref{epsimag}).
The $\mathbf{k}$-point mesh used in an ordinary $GW$ calculation is not sufficiently dense to
obtain an accurate dielectric function, however, and it is computationally very expensive to
increase the density of this mesh. As we will show below, the main difference between the $GW$ and
the GGA energies for the systems studied, is the size of the band gap between the occupied and the
unoccupied states. The differences between the dispersions of the $GW$ and GGA bands are relatively
small. Therefore we adopt a simple ``scissors" operator approximation for the energy differences in
Eq.~(\ref{epsimag}),\cite{delsole}
\begin{equation}
\label{scissors} \epsilon_{c\bk+\q}-\epsilon_{v\bk}=\epsilon_{c\bk+\q}^{GGA}-\epsilon_{v\bk}^{GGA}+
E_{gap}^{GW}-E_{gap}^{GGA}.
\end{equation}


\section{Results}

\subsection{Simple hydrides}

\subsubsection{Test calculations}\label{testcalc}

Calculations on $\lih$ and $\nah$ are relatively straightforward because these compounds have a
simple rocksalt structure. They can be used to benchmark the calculations. The calculated single
particle band gaps of $\lih$ and $\nah$ are listed in Table~\ref{simplegaps}. As is usual, DFT
severely underestimates the gap, with LDA giving smaller values than GGA. Our calculated $GW$ gaps
for $\lih$ and $\nah$ are close to those obtained in recent PAW all-electron $GW$
calculations.\cite{lebegue1,lebegue2} As stated in the previous section, our calculations use
pseudopotentials and take into account explicitly the valence electrons only. The $GW$ band gaps as
calculated from the QP energies obtained by solving Eq.~(\ref{dec_eq}), are indicated by $E_g^{GW}$
in Table~\ref{simplegaps}. They are somewhat larger than the PAW values. If one applies the core
correction of Eq.~(\ref{vxc_eq}), our $GW$ band gaps become somewhat smaller than the PAW values,
as shown by the column marked $E_g^{GW,\mathrm{core}}$ in Table~\ref{simplegaps}. The differences
between the pseudopotential and the PAW $GW$ gaps are small, however, i.e. of the order of 2-4\%
and the PAW values are in between the $E_g^{GW,\mathrm{core}}$ and $E_g^{GW}$ values.

\begin{table}[!tbp]
\begin{ruledtabular}
\caption{\label{simplegaps} Single particle band gaps $E_g$ (eV) of simple hydrides from DFT (GGA
and LDA) and $GW$ calculations. $E_g^{GW,\mathrm{core}}$ refers to applying the correction of
Eq.~(\ref{vxc_eq}).}
\begin{tabular}{lcccccc}
   & $E_g^{GGA}$ & $E_g^{LDA}$ & $E_g^{GW}$ & $E_g^{GW,\mathrm{core}}$ & $GW$ lit. \\
\hline $\lih$  & 3.00 & 2.61 & 4.75 & 4.54
&4.64\footnotemark[1], 5.24\footnotemark[2], 5.37\footnotemark[3]\\
&&&&& 4.99$^{\mathrm{exp},}$\footnotemark[4]\\
$\nah$  & 3.79 & 3.42 & 5.87 & 5.50 &5.68\footnotemark[5]\\
$\mgh$  & 3.79 & 3.36 & 5.64 & 5.32 &5.25\footnotemark[6], 5.58\footnotemark[7]\\
$\alh$  & 2.18 & 1.79 & 4.31 & 3.54 &\\
\end{tabular}
\end{ruledtabular}
\footnotetext[1]{Ref. \onlinecite{lebegue1}} \footnotetext[2]{COHSEX, Ref.~\onlinecite{baroni}}
\footnotetext[3]{Ref. \onlinecite{shirley}} \footnotetext[4]{Experimental gap at $T=4.2$~K, Ref.
\onlinecite{plek}} \footnotetext[5]{Ref. \onlinecite{lebegue2}} \footnotetext[6]{cited in Ref.
\onlinecite{isidorsson}} \footnotetext[7]{Ref. \onlinecite{alouani}}
\end{table}

The band gap of $\lih$ given in Ref.~\onlinecite{baroni} has been calculated using the rather crude
COSHEX approximation, which is known to lead to a much higher gap.\cite{lebegue1} The value given
in Ref.~\onlinecite{shirley} is much higher than that obtained in other $GW$ calculations,
including ours, for which we have no explanation. The experimental band gap of $\lih$ is higher
than the calculated $GW$ values,\cite{plek} but the difference is within the usual
5-10\%.\cite{aryasetiawan,rubio} To our knowledge no experimental data are available for $\nah$.

The band gap of $\mgh$ is calculated for the optimized rutile or
$\alpha$-structure,\cite{vansetten} which is the most stable structure at room temperature and
ambient pressure.\cite{vajeeston} The $E_g^{GW,\mathrm{core}}$ value we obtain is very close to
that cited in Ref.~\onlinecite{isidorsson}. A recently obtained PAW value for the band gap of
$\mgh$ is again in between our $E_g^{GW,\mathrm{core}}$ and $E_g^{GW}$ values.\cite{alouani} For
$\alh$ no other data are available to our knowledge. One can observe that core effects are
relatively large in this compound.

In the following we will use the $E_g^{GW,\mathrm{core}}$ values. The validity of using the
scissors operator approximation for calculating the optical response, see Eq.~\ref{scissors}, is
illustrated by comparing band widths calculated with DFT and $GW$. The valence band widths are
given in Table~\ref{bandwidths}. The difference between the $GW$ and the LDA values is within 3\%
and the difference between the $GW$ and the GGA values is within 10\%. Note that the latter is on
the same scale as the difference between the LDA and the GGA values. This accuracy is acceptable
for our purposes. We have also checked in more detail that the dispersions of the individual bands
in the DFT and $GW$ band structures are very similar.

\begin{table}[!tbp]
\caption{\label{bandwidths}Valence band widths (eV) from DFT (GGA and LDA) and $GW$ calculations.
The directionally averaged real part of the static dielectric constant, calculated with ($LFE$) and
without local field effects.}
\begin{ruledtabular}
\begin{tabular}{lcccccc}
   & GGA & LDA & $GW$ & $\varepsilon^\infty_{LFE}$ & $\varepsilon^\infty$\\
\hline
$\lih$  & 5.41 & 5.62 & 5.81 & 4.28 & 4.34\\
$\nah$  & 3.58 & 4.00 & 3.99 & 3.03 & 3.06\\
$\mgh$  & 6.34 & 6.62 & 6.66 & 3.90 & 3.98\\
$\alh$  & 8.60 & 8.82 & 8.92 & 4.43 & 4.55\\
\end{tabular}
\end{ruledtabular}
\end{table}

Table~\ref{bandwidths} also lists the static dielectric constant calculated with and without local
field effects. The small differences between these numbers indicate that it is reasonable to
neglect local field effects in calculating the dielectric function.

\subsubsection{$\lih$ and $\nah$}\label{lihnah}

The calculated optimized lattice parameters of $\lih$ and $\nah$ in the rock salt structure are
4.02 and 4.83~\AA, respectively. These values are somewhat smaller than the experimental lattice
parameters of 4.09 and 4.91~\AA\ due to neglecting the zero point motions of the hydrogen atoms, as
discussed in Ref.~\onlinecite{martins}.

\begin{figure}[!tbp]
\includegraphics[angle=270, width=7.7cm]{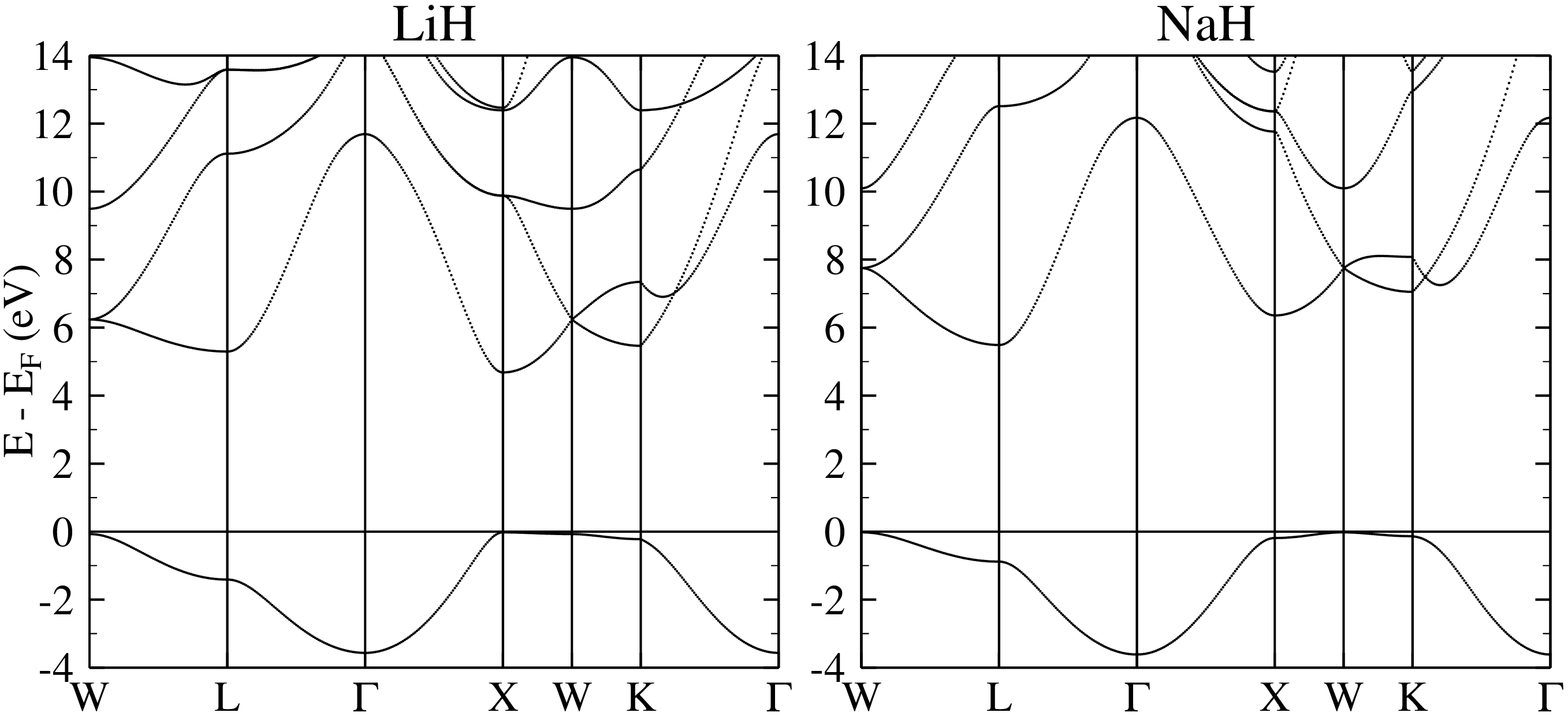}
\includegraphics[angle=270, width=9.0cm]{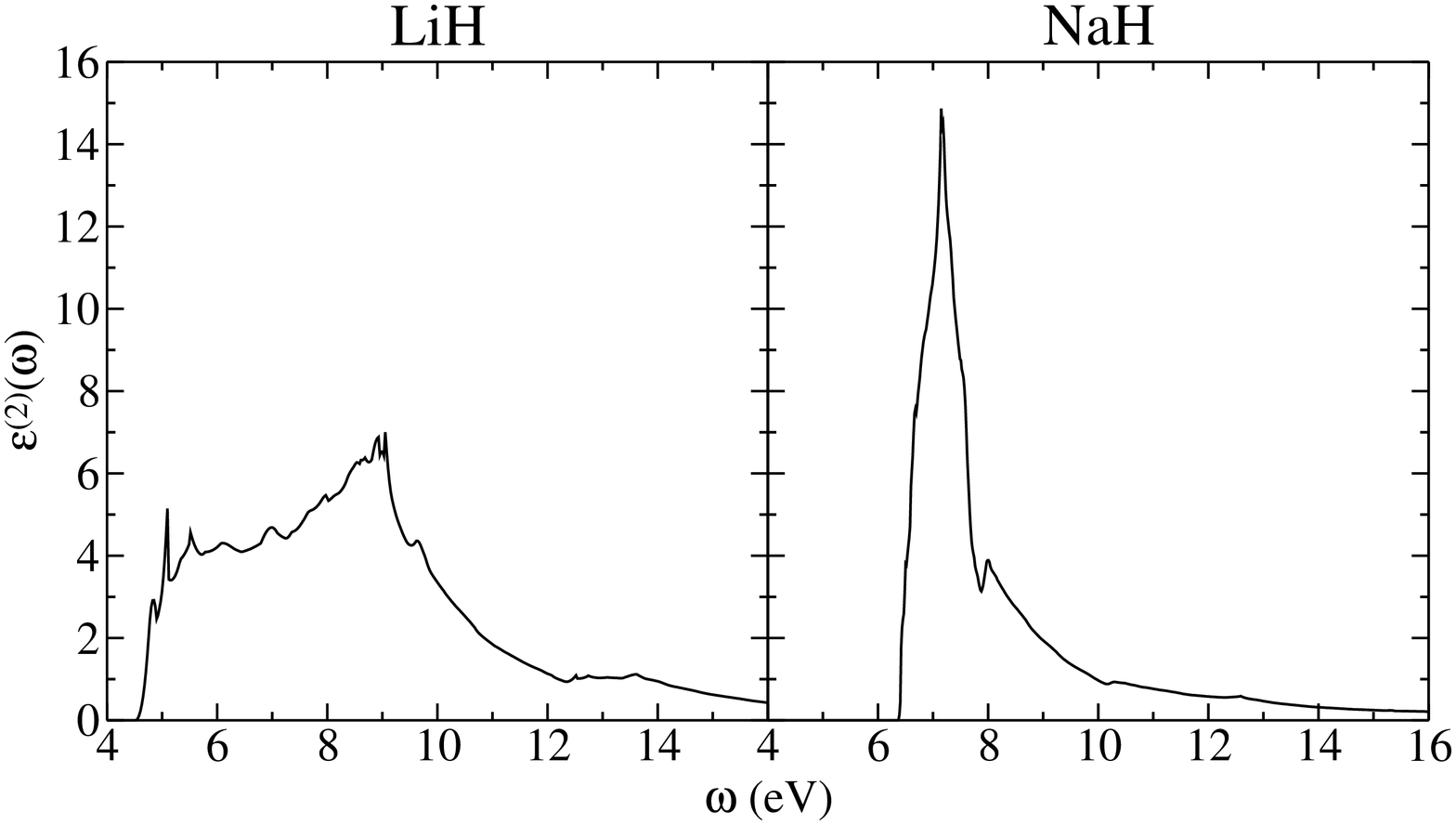}
\caption{\label{lina}Upper panels: electronic band structures of $\lih$ and $\nah$. The zero of the
energy scale is at the top of the valence band. Lower panels: imaginary parts of the directionally
averaged macroscopic dielectric functions of $\lih$ and $\nah$. The calculated optical gaps of
$\lih$ and $\nah$ are 4.54~and 6.37~eV, respectively. Unless explicitly stated otherwise, the
results presented in the figures are based upon GGA calculations modified by a scissors operator
extracted from the $GW$ results.}
\end{figure}

The band structures and the directionally averaged $\varepsilon^{(2)}(\omega)$ of $\lih$ and $\nah$
are shown in Fig.~\ref{lina}. The valence bands in both $\lih$ and $\nah$ are strongly dominated by
hydrogen, which reflects the ionic character of the bonding.\cite{lebegue1,baroni,shirley,lebegue2}
The conduction bands have a mixed hydrogen and cation character. The lattice parameter of $\lih$ is
significantly smaller than that of $\nah$. Because of the smaller distance between the hydrogen
atoms the band widths in $\lih$ are generally larger than in $\nah$. This is most easily observed
in the valence band, whose dispersion is quite similar in $\lih$ and $\nah$, but the ($GW$) valence
band width in $\lih$ is 5.81~eV, whereas in $\nah$ it is 3.99~eV.

The conduction bands of the two compounds are qualitatively different. In $\lih$ the conduction
band minimum is at X, whereas in $\nah$ it is at L, which causes the gap in $\lih$ to be direct,
whereas in $\nah$ it is indirect. In $\nah$ there is little participation of cation states in the
lower lying conduction bands. A calculation on an fcc lattice of H$^-$ ions with the $\nah$ lattice
parameter and a homogeneous background charge instead of the Na$^+$ cations, gives essentially the
same band structure. In $\lih$ the participation of the Li cations to the conduction band is
larger. The conduction band minimum at X is lowered in energy because here the Li-$2p$ state
participates in a bonding combination with hydrogen states. Similarly, the conduction band minimum
at L is raised in energy because the Li-$2s$ state contributes to an anti-bonding combination with
hydrogen states.

The differences between the band structures of $\lih$ and $\nah$ lead to markedly different
dielectric functions, as is shown in the lower panel of Fig.~\ref{lina}. The calculated direct
optical gap in $\lih$ is 4.54~eV, whereas in $\nah$ it is a much larger~6.37~eV. The main peak in
the $\nah$ spectrum originates from transitions at L and X, whose onset is at comparable photon
energies. Both valence and conduction bands of $\nah$ have hydrogen character and the oscillator
strength of these transitions is rather large. The result is a sharp and dominant peak just above
the onset of the optically allowed transitions.

The optical spectrum of $\lih$ has more structure. The onset at 4.54~eV is due to transitions at X,
and at a somewhat higher energy transitions near K and W contribute. At energies $\sim 9$~eV
transitions at L become allowed, which results in a peak in the dielectric function at that energy.
Since the bands of $\lih$ have a larger dispersion than those of $\nah$, the optical response of
$\lih$ is spread out over a larger energy range.

\subsubsection{$\mgh$}

$\alpha$-$\mgh$ has the rutile structure, i.e.\ space group $P4_2/mnm$ with Mg, H atoms in $2a$,
$4f$ Wyckoff positions, respectively, and two formula units per unit cell. The optimized calculated
lattice parameters are $a=4.52$~\AA\ and $c=3.01$~\AA, with the H atoms at $x=0.304$. This is in
good agreement with the experimental values $a=4.50$~\AA, $c=3.01$~\AA\ and $x=0.304$.\cite{bortz}
The magnesium atoms are sixfold octahedrally coordinated by the hydrogen atoms at distances between
1.94 and 1.96~\AA. Each MgH$_6$ octahedron shares the hydrogen atoms at its corners with
neighboring octahedra. Each hydrogen atom is shared by three octahedra and is therefore
coordinated by three magnesium atoms.

\begin{figure}[!tbp]
\includegraphics[angle=270, width=7.7cm]{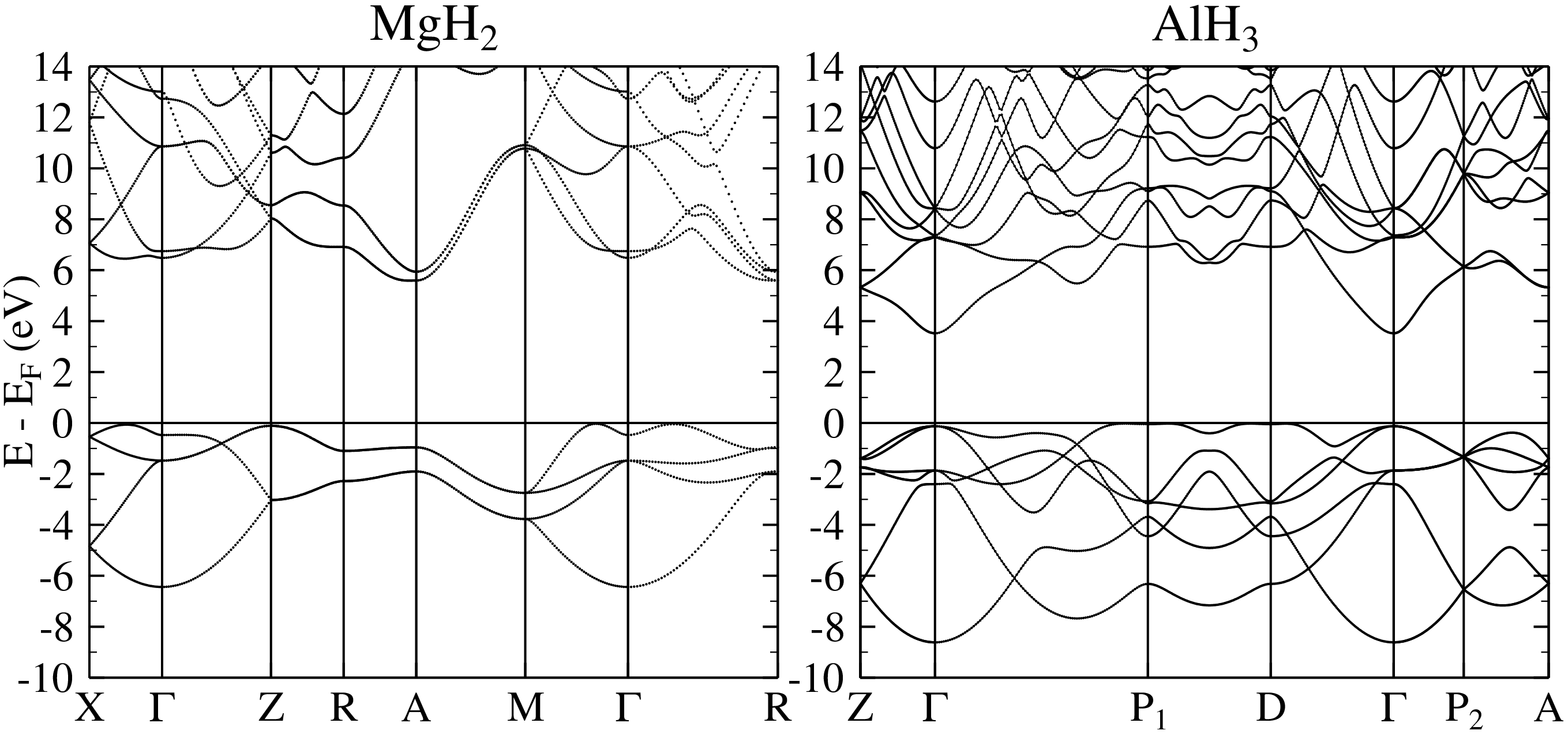}
\includegraphics[angle=270, width=9cm]{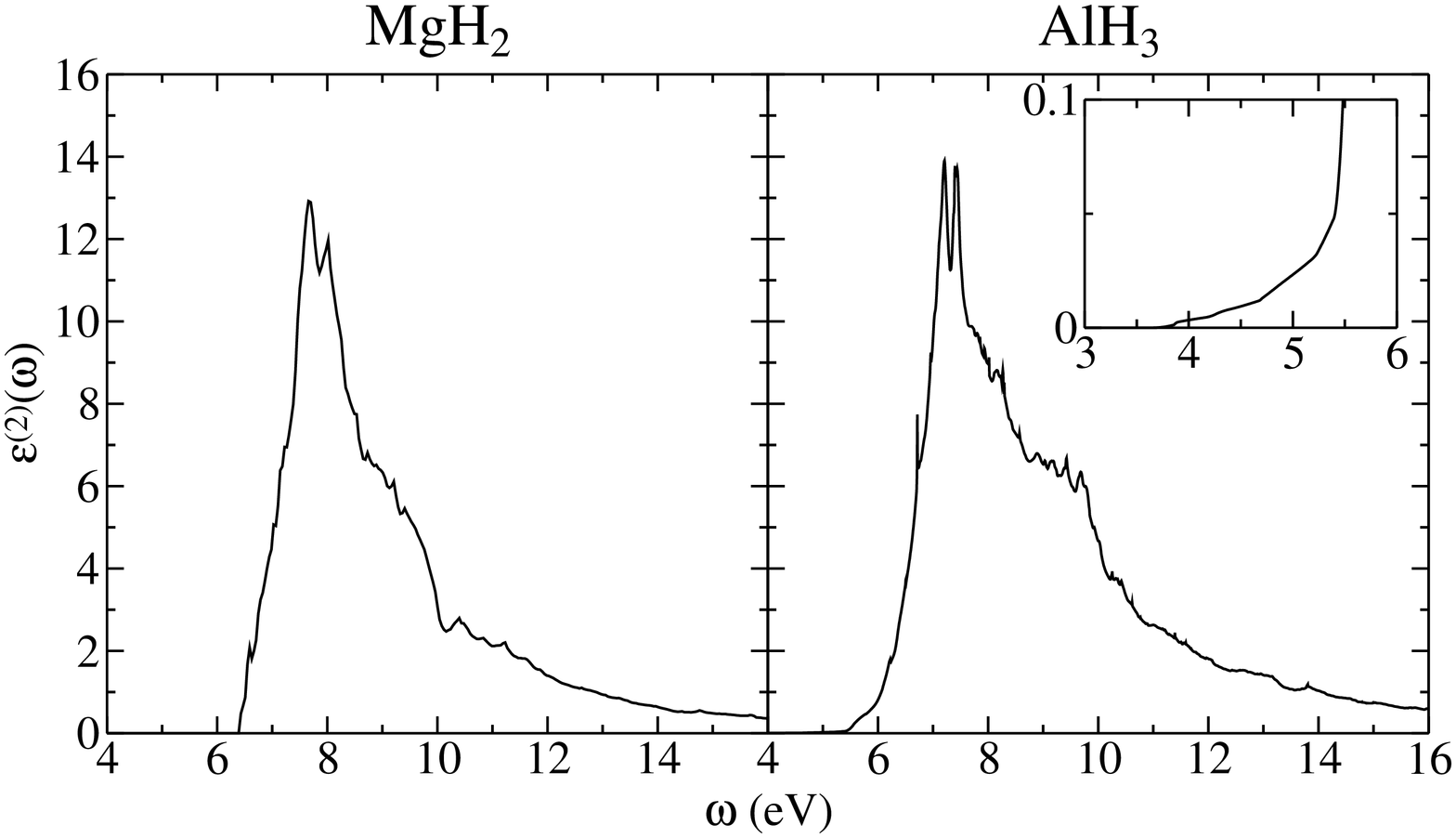}
\caption{\label{mgal}Upper panels: electronic band structures of $\mgh$ and $\alh$. The zero of the
energy scale is at the top of the valence band. P$_1$ and P$_2$ correspond to the points
(0.5,-1,0.5) and (0.5,0.5,0.5). Lower panels: imaginary parts of the directionally averaged
macroscopic dielectric functions of $\mgh$ and $\alh$. The calculated optical gaps of $\mgh$ and
$\alh$ are 6.19 and 3.54~eV, respectively.}
\end{figure}

The band structure of $\mgh$ is shown in Fig.~\ref{mgal}. Our results are in general agreement with
those obtained in previous work.\cite{rici,alouani} There is a small hybridization between the H
and the Mg states in the valence bands, with the lowest two valence bands having some Mg $s$ and
the highest two having some Mg $p$ character, respectively. As in the case of $\lih$ and $\nah$,
the valence bands have a dominant hydrogen character, however. The conduction bands have a mixed
magnesium hydrogen character and the bottom of the conduction band has a considerable Mg $3s$
contribution. $\mgh$ has a calculated indirect gap of 5.32~eV, see Table~\ref{simplegaps}, whereas
the direct optical gap is 6.19~eV. The experimental optical gap obtained in
Ref.~\onlinecite{isidorsson} is $5.6 \pm 0.1$~eV, which would indicate that our $GW$ result
overestimates the gap by 10\%.

The lower panel of Fig.~\ref{mgal} shows the dielectric function of $\mgh$. The onset of optical
transitions occurs almost at the same energy at various regions throughout the Brillouin zone,
which results in a steep rise of the dielectric function and a peak close to 8~eV. At energies
above 9~eV transitions from the lower valence bands start to play a role, see for instance the
interval Z-M in the upper panel of Fig.~\ref{mgal}. This results in a shoulder in the dielectric
function at $\sim 9$~eV. Finally, the shoulder at $\sim 11$~eV in the spectrum involves transitions
to higher lying conduction bands, associated with rather delocalized states having considerable Mg
character.


\subsubsection{$\alh$}\label{alh3}

$\alh$ has a rhombohedral structure with space group $R\overline{3}c$ and the Al, H atoms in the
$6b$, $18e$ Wyckoff positions, respectively. The unit cell contains two formula units. The
optimized lattice parameters are $a=4.49$~\AA\ and $c=11.82$~\AA, with the H atoms at $x=0.623$.
These values are in good agreement with the experimental values $a=4.45$~\AA, $c=11.80$~\AA\ and
$x=0.628$.\cite{turley} The interatomic Al-Al distances in the $ab$ plane and along the $c$ axis
are 4.45~\AA\ and 3.24~\AA, respectively. The aluminium atoms form a distorted face centered
structure, where each Al atom is octahedrally coordinated by H atoms with an Al-H distance of
1.75~\AA. Each AlH$_6$ octahedron shares its corners with neighboring octahedra and each H atom at
a corner forms a bridge between two Al atoms. Since these bridges are not linear, i.e. the Al-H-Al
bond angle is $141^\mathrm{o}$, the octahedra are tilted with respect to one another.

The band structure of $\alh$ is shown in Fig.~\ref{mgal}. There is hybridization between H and Al
states, but the six valence bands are dominated by hydrogen, as are the valence bands of the other
hydrides. In contrast to $\mgh$, $\alh$ has a direct band gap, which is located at $\Gamma$. The
band gap is 3.54~eV, which is notably smaller than the gap in the other hydrides discussed above.
This is caused by a single conduction band that disperses to $\sim$~2~eV below the other conduction
bands. This band has a large Al $3s$ contribution.

The dielectric function of $\alh$ is shown in the lower panel of Fig.~\ref{mgal}. Although the
optical response starts at the direct gap of 3.54~eV, see the inset in Fig.~\ref{mgal}, it reaches
significant values only above 6~eV. The weak response between 3.54 and 6~eV is caused by the fact
that only a single conduction band contributes with a low density of states. Moreover, since that
band has Al $3s$ character, whereas the valence bands have dominant H character, the oscillator
strength of these transitions is small. The dielectric function rises sharply above 6~eV and peaks
above 7~eV. The spectrum has a distinct broad shoulder between 9 and 10~eV and also some weaker
shoulders at higher energies.

\begin{figure}[!tbp]
\includegraphics[width=8cm]{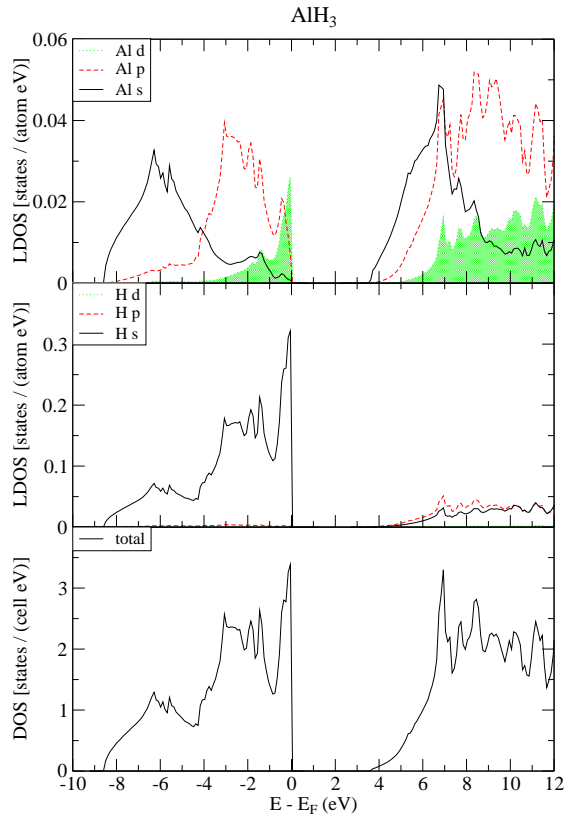}
\caption{\label{dosalh3}(Color online) Local densities of states (LDOS) in atomic angular momentum
projection; bottom panel: total density of states. The Fermi energy is at the top of the valence
band.}
\end{figure}

In order to interpret the dielectric function it is instructive to analyze the total density of
states (DOS) and the local density of states (LDOS) of $\alh$. These are shown in
Fig.~\ref{dosalh3}.\@
The valence DOS has a sharp peak just below the Fermi level and two broad peaks at $\sim -2.5$~eV
and $\sim -6$~eV below the Fermi level. These three peaks originate from respectively $3s$, $3p$
and $3d$ aluminium states hybridizing with the $1s$ hydrogen states, as can be observed in the
upper two panels of Fig.~\ref{dosalh3}. In the solid the states are broadened into strongly
overlapping bands. In the dielectric function transitions from the highest two of these valence
peaks gives rise to the structure between 6 and 10~eV in Fig.~\ref{mgal}. The energy associated
with transitions from the lowest valence peak is too high to give any significant contribution to
the dielectric function.

Qualitatively the optical spectra of $\alh$ and $\mgh$ show some similarity, despite the difference
in structure between these materials. In both cases the dielectric function sharply rises above 6
eV and peaks between 7 and 8~eV. The spectrum of $\alh$ is broader due to a larger dispersion of
the bands, reflecting the somewhat denser packing of the hydrogen atoms in this compound. From
Fig.~\ref{lina} one observes that the dielectric function of $\nah$ also rises steeply above 6~eV
and peaks just above 7~eV. The spectrum of $\nah$ is narrower than that of $\mgh$ and $\alh$,
reflecting the less dense packing of hydrogen atoms in this compound, which results in a smaller
band dispersion. Only the spectrum of $\lih$ is qualitatively different as it rises below 5~eV in a
broad shoulder. As discussed in Sec.~\ref{lihnah}, there is a significant contribution from the Li
states in this case.

\subsection{The binary hydrides $\litt$ and $\naa$}\label{octalabel}

The optimized atomic positions of $\litt$ and $\naa$ are given in Table~\ref{pos}. Both $\litt$ and
$\naa$ consist of a stacking of AlH$_6$ octahedra and alkali cations. The octahedra are slightly
distorted with Al-H distances of 1.75~\AA\ in $\litt$ and 1.78 to 1.80~\AA\ in $\naa$. These
compounds contain a relatively large fraction of alkali cations. Since sodium atoms are larger than
lithium atoms, the distance between the AlH$_6$ octahedra in $\naa$ is considerably larger. The
Al-Al distance in $\naa$ is 5.59~\AA, whereas in $\litt$ it is 4.88~\AA. As for the simple hydrides
discussed in Sec.~\ref{lihnah}, this size effect leads to noticeable differences in the electronic
structure and the optical properties of $\litt$ and $\naa$.

\begin{table}[!tbp]
\caption{\label{pos} Optimized atomic positions in the binary hydrides. The labels ``1a'' etc.
refer to the Wyckoff positions. The cell parameters are taken from the references. The structures
are in good agreement with previous experimental and theoretical
work.\cite{chou,lov,ron,chun,fich,arr,agu,vaj,hau,song06}}
\begin{ruledtabular}
\begin{tabular}{llllccc}
Compound &Space group &        &    &$x$&$y$&$z$\\
         &unit cell   &        &    &   &   &   \\
\hline
$\litt$ & $R\bar{3}$ (148)\footnotemark[1] & 6f&Li&0.9329&0.4396&0.7512\\
        & $a$ =  5.64 \AA                  & 1a&Al&0&0&0\\
        & $\alpha$ = 91.4$^\circ$          & 1b&Al&1/2&1/2&1/2\\
        &                                  & 6f&H&0.7054&0.9287&0.0675\\
        &                                  & 6f&H&0.7941&0.5885&0.4518\\
\hline
$\naa$ & $P2_1/n$ (14)\footnotemark[2]     & 2b&Na&0&0&1/2\\
       & $a$ = 5.51 \AA                    & 4e&Na&0.9908&0.4566&0.2553\\
       & $b$ = 5.67 \AA                    & 2a&Al&0&0&0\\
       & $c$ = 7.91 \AA                    & 4e&H&0.0983&0.0515&0.2125\\
       & $\beta$ = 89.9$^\circ$            & 4e&H&0.2331&0.3327&0.5413\\
       &                                   & 4e&H&0.1583&0.2622&0.9347\\
\hline
$\lit$ & $P2_1/c$ (14)\footnotemark[3]     & 4e&Li&0.5727&0.4650&0.8254\\
       & $a$ = 4.84 \AA                    & 4e&Al&0.1395&0.2016&0.9314\\
       & $b$ = 7.81 \AA                    & 4e&H&0.1784&0.0988&0.7614\\
       & $c$ = 7.83 \AA                    & 4e&H&0.3561&0.3720&0.9775\\
       & $\beta$ = 112.1$^\circ$           & 4e&H&0.2394&0.0816&0.1142\\
       &                                   & 4e&H&0.7953&0.2631&0.8714\\
\hline
$\na$  & $I4_1/a$ (88)\footnotemark[4]     & 4b&Na&0&1/4&5/8\\
       & $a$ = 5.01 \AA                    & 4a&Al&0&1/4&1/8\\
       & $c$ = 11.31 \AA                   & 16f&H&0.2354&0.3900&0.5454\\
\hline
$\mg$  & $P\bar{3}m1$ (164)\footnotemark[5]& 1a& Mg&0&0&0\\
       & $a$ = 5.23 \AA                    & 2d&Al&1/3&2/3&0.7064\\
       & $c$ = 6.04 \AA                    & 2d&H&1/3&2/3&0.4415\\
       &                                   & 6i&H&0.1680&-0.1680&0.8118\\
\end{tabular}
\end{ruledtabular}
\footnotetext[1]{Ref.~\onlinecite{bri}} \footnotetext[2]{Ref.~\onlinecite{chou}}
\footnotetext[3]{Ref.~\onlinecite{lov}} \footnotetext[4]{Ref.~\onlinecite{hau2}}
\footnotetext[5]{Ref.~\onlinecite{vansetten}}
\end{table}

\begin{figure*}[!tbp]
\includegraphics[angle=270,width=16cm]{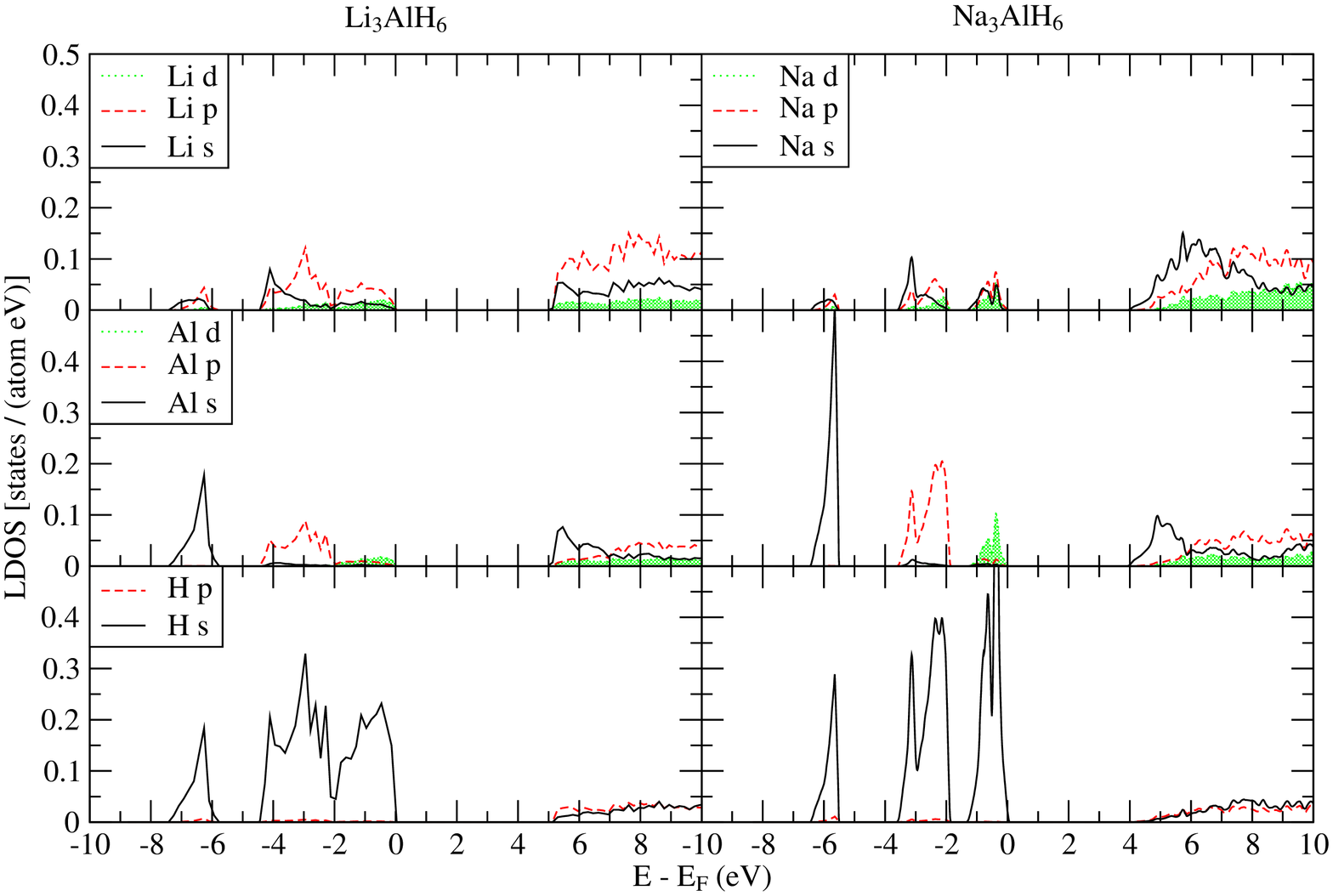}
\caption{\label{doslina}(Color online) Local densities of states (LDOS) of $\litt$ and
$\naa$.\cite{AR} For clarity the area under the $d$-line is shaded. The Fermi energies are at the
top of the valence band. The conduction band DOS is almost constant up to at least 22~eV.}
\end{figure*}

The local electronic densities of states (LDOS) of both compounds is given in Fig.~\ref{doslina}.\@
As in AlH$_3$ the valence bands have dominant hydrogen character, although there is Al character
mixed in. The splitting into three peaks with approximate relative intensity 1:3:2 can be
interpreted in terms of an octahedral ligand field splitting.\cite{chou} The peaks correspond
respectively to the $3s$, $3p$ and $3d(e_g)$ aluminum states hybridizing with the $1s$ hydrogen
levels in the AlH$_6$ octahedra. As can be observed in Fig.~\ref{doslina} the $spd$ splitting in
$\litt$ and $\naa$ is comparable, which reflects the similarity of the AlH$_6$ octahedral structure
in both compounds. Comparing to Fig.~\ref{dosalh3} one observes that the splitting is also
comparable to that in AlH$_3$, again suggesting the similarity in the octahedral structure.

The interaction between the octahedra in the solid results in a broadening of the three peaks.
Unlike in AlH$_3$ the AlH$_6$ octahedra in $\litt$ and $\naa$ are not directly connected, which
limits the interaction and the broadening. Therefore, the three peaks remain non-overlapping. One
expects their widths to increase as the distance between the octahedra decreases and indeed the
valence peaks in the DOS of $\litt$ are wider than in $\naa$. The conduction band of both compounds
is rather featureless up to at least 22 eV. There is somewhat more cation $s$ character at the
bottom of the conduction band in $\naa$, whereas the bonding in $\litt$ probably has a somewhat
more covalent character, as in the simple hydrides.

The calculated band gaps are given in Table~\ref{bingaps}. As discussed in Sec.~\ref{testcalc}, DFT
calculations severely underestimate the gap, with LDA giving a $\sim 0.5$~eV smaller value than
GGA. The most important results are in the last two columns of Table~\ref{bingaps}, which give the
$GW$ single particle gap and the direct optical gap. The single particle gap is indirect in $\litt$
and direct in $\naa$.\cite{na3_remark} The fairly large difference between the gaps of $\litt$ and
$\naa$ is striking. Moreover, the fact that the gap of $\litt$ is larger is somewhat
counterintuitive. Naively one would expect that the larger broadening of the AlH$_6$ octahedron
levels discussed above would narrow the gap, since it leads to a larger valence band width. The
origin of the band gap difference between $\litt$ and $\naa$ is discussed in Sec.~\ref{discussion}.

\begin{table}[!tbp]
\begin{ruledtabular}
\caption{\label{bingaps} Single particle band gaps $E_g$ (eV) of the binary hydrides calculated
with DFT (GGA and LDA) and $GW$. $E_g^{GW,\mathrm{core}}$ refers to applying the correction of
Eq.~(\ref{vxc_eq}); opt$^{GW}$ refers to the direct optical gap}
\begin{tabular}{lcccc}
   & $E_g^{GGA}$ & $E_g^{LDA}$ & $E_g^{GW,\mathrm{core}}$ & opt$^{GW}$ \\
\hline
$\litt$  & 3.65 & 3.13 & 5.10 & 5.31\\
$\naa$   & 2.54 & 2.00 & 3.94 & 3.94\\
$\lit$   & 4.67 & 4.19 & 6.55 & 6.89\\
$\na$    & 4.63 & 4.12 & 6.41 & 6.50\\
$\mg$    & 4.40 & 3.99 & 6.48 & 6.87\\
\end{tabular}
\end{ruledtabular}
\end{table}

\begin{figure}[!tbp]
\includegraphics[angle=270, width=9cm]{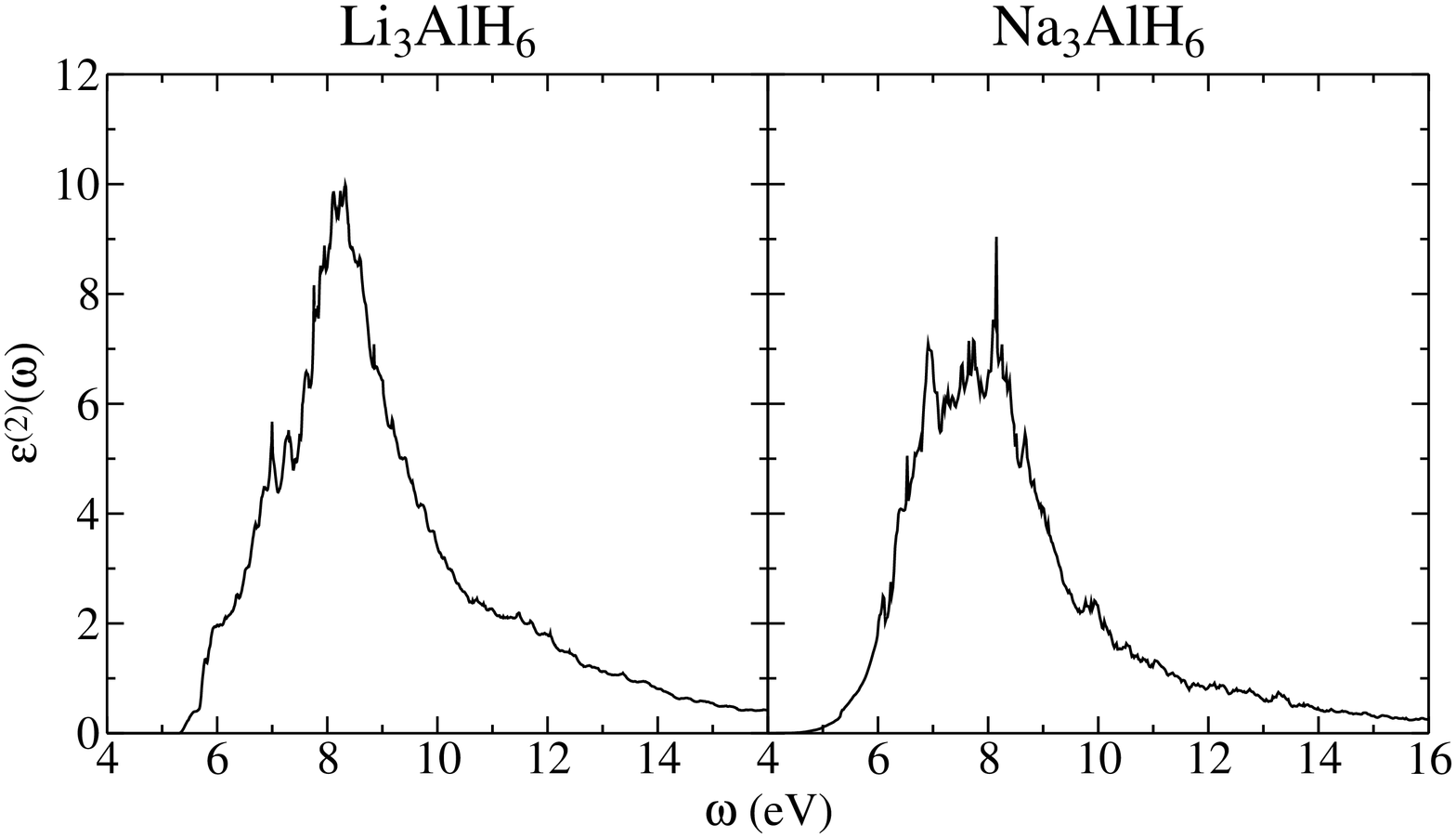}
\caption{\label{diefli3na3}Imaginary parts of the directionally averaged macroscopic dielectric
functions of $\litt$ and $\naa$.}
\end{figure}

The directionally averaged dielectric functions of $\litt$ and $\naa$ are shown in
Fig.~\ref{diefli3na3}.\@
The dielectric function of $\litt$ has a shoulder starting above 6~eV, a peak just above 8~eV and a
shoulder below 12~eV. Since the conduction band DOS is rather uniform and featureless up to at
least 22~eV, these features in the dielectric function can be directly linked to transitions from
the three octahedron valence peaks.

Despite the much smaller band gap of $\naa$ the dielectric function starts to increase appreciably
only at an energy between 5 and 6~eV, which is not that much lower than in $\litt$. Transitions
from the top two valence peaks give rise to the complicated pattern between 6 and 9~eV; transitions
from the third valence peak gives the above 10~eV. Qualitatively these spectra have a resemblance
to that of AlH$_3$, see Fig.~\ref{mgal}, reflecting the dominant role played by the AlH$_6$
octahedra.

\subsection{The binary hydrides $\lit$, $\na$ and $\mg$}
\label{tetralabel}

The optimized structures of $\lit$, $\na$ and $\mg$ are given in Table~\ref{pos}. For $\lit$ and
$\na$ we have used the experimental unit cells and optimized the atomic positions only; for $\mg$
we have also optimized the size and shape of the unit cell.\cite{vansetten} All three materials
consist of a packing of AlH$_4$ tetrahedra and alkali or alkaline earth cations. The tetrahedra are
slightly distorted  and the Al-H distances vary from 1.62 to 1.65~\Am in $\lit$, 1.64~\Am in $\na$
and from 1.60 to 1.62~\Am in $\mg$. Unlike the two compounds discussed in the previous section, the
volume fraction taken up by the cations is relatively small and the distance between the AlH$_4$
tetrahedra is hardly influenced by the size of the cations. The Al-Al distance is 3.75~\Am in
$\lit$, 3.78~\Am in $\na$  and 3.86~\Am in $\mg$.

\begin{figure*}[!tbp]
\includegraphics[angle=270,width=16cm]{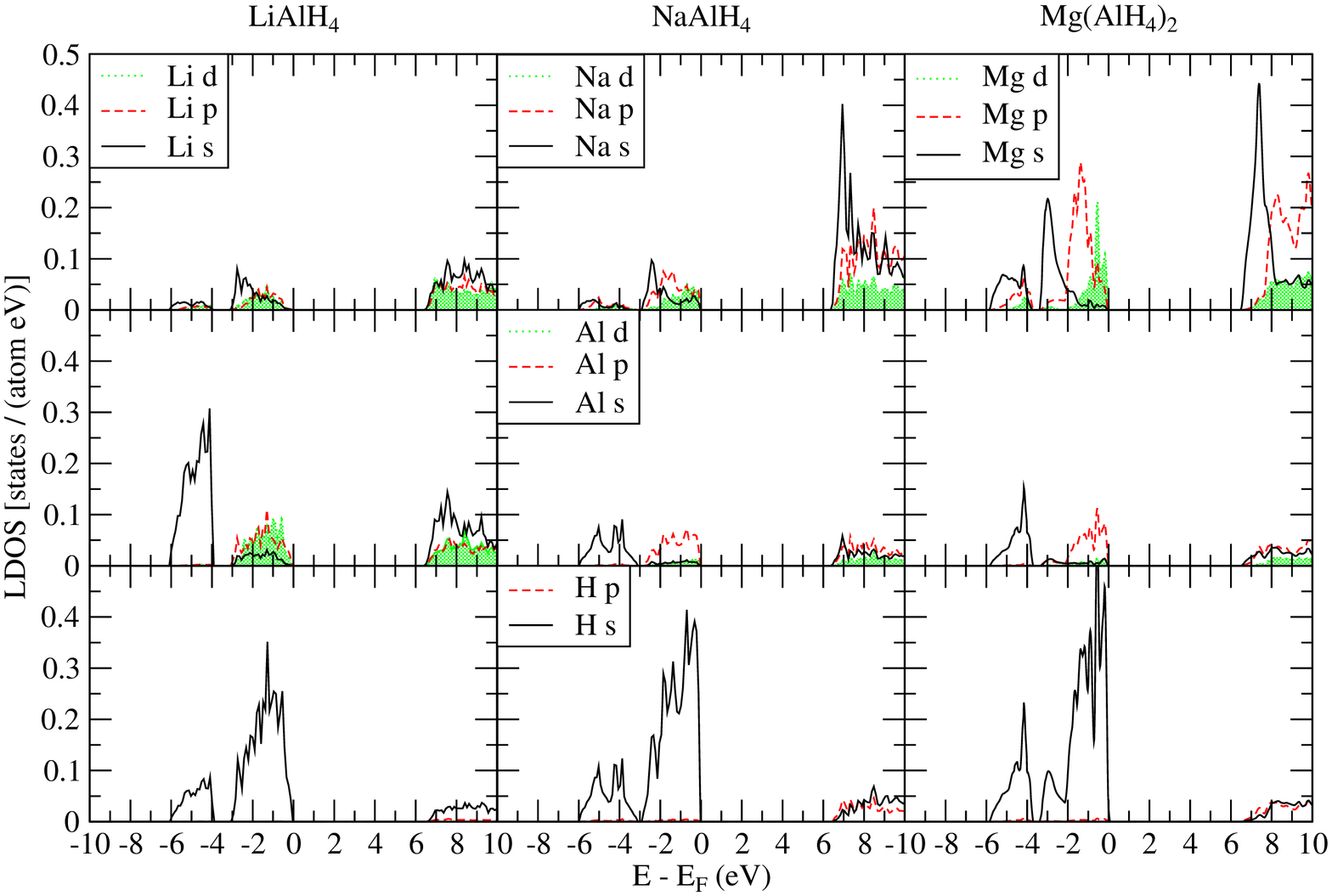}
\caption{\label{doslinamg}(Color online) Local densities of states (LDOS) of $\lit$, $\na$ and
$\mg$.\cite{AR} For clarity the area under the $d$-line is shaded. The Fermi energies are at the
top op of the valence band. The conduction band DOS is almost constant up to at least 22~eV.}
\end{figure*}

The LDOS of $\lit$, $\na$ and $\mg$ is shown in Fig.~\ref{doslinamg}.\@
As for the compounds discussed before, the valence bands have dominant hydrogen character with some
Al character mixed in. The splitting into two peaks with approximate relative intensity 1:3 is due
to a tetrahedral ligand field splitting of the Al $3p$ and $3s$ levels hybridized with H $1s$
levels in AlH$_4$.\cite{chou} The splitting is comparable in all three compounds, which reflects
the similarity of the structure and bonding of the AlH$_4$ tetrahedra in these compounds. The
interaction between the tetrahedra causes a broadening of these levels in the solid.
Fig.~\ref{doslinamg} shows that also the resulting band widths of these valence states is
comparable in all three compounds. Apparently the widths are not extremely sensitive to the details
of the structure, which are quite different for $\lit$, $\na$ and $\mg$. They
are sensitive to the distance between the tetrahedra, but this is comparable for the three
compounds.

Compared to the valence bands, the features in the conduction bands are less distinct. Both in
$\na$ and in $\mg$ the bottom of the conduction band has considerable $s$ character derived from
the empty $3s$ state of the cation. The conduction band in $\lit$ is featureless. The LDOS on the
Al atoms is very similar in $\na$ and in $\mg$, but there are small differences with $\lit$. There
is significant Al $d$ character in the valence band in the latter compound, and almost none in the
other two compounds. In the conduction band of $\lit$ there is a considerable Al $s$
contribution, and much less in the other compounds.

\begin{figure}[!tbp]
\includegraphics[angle=270, width=9cm]{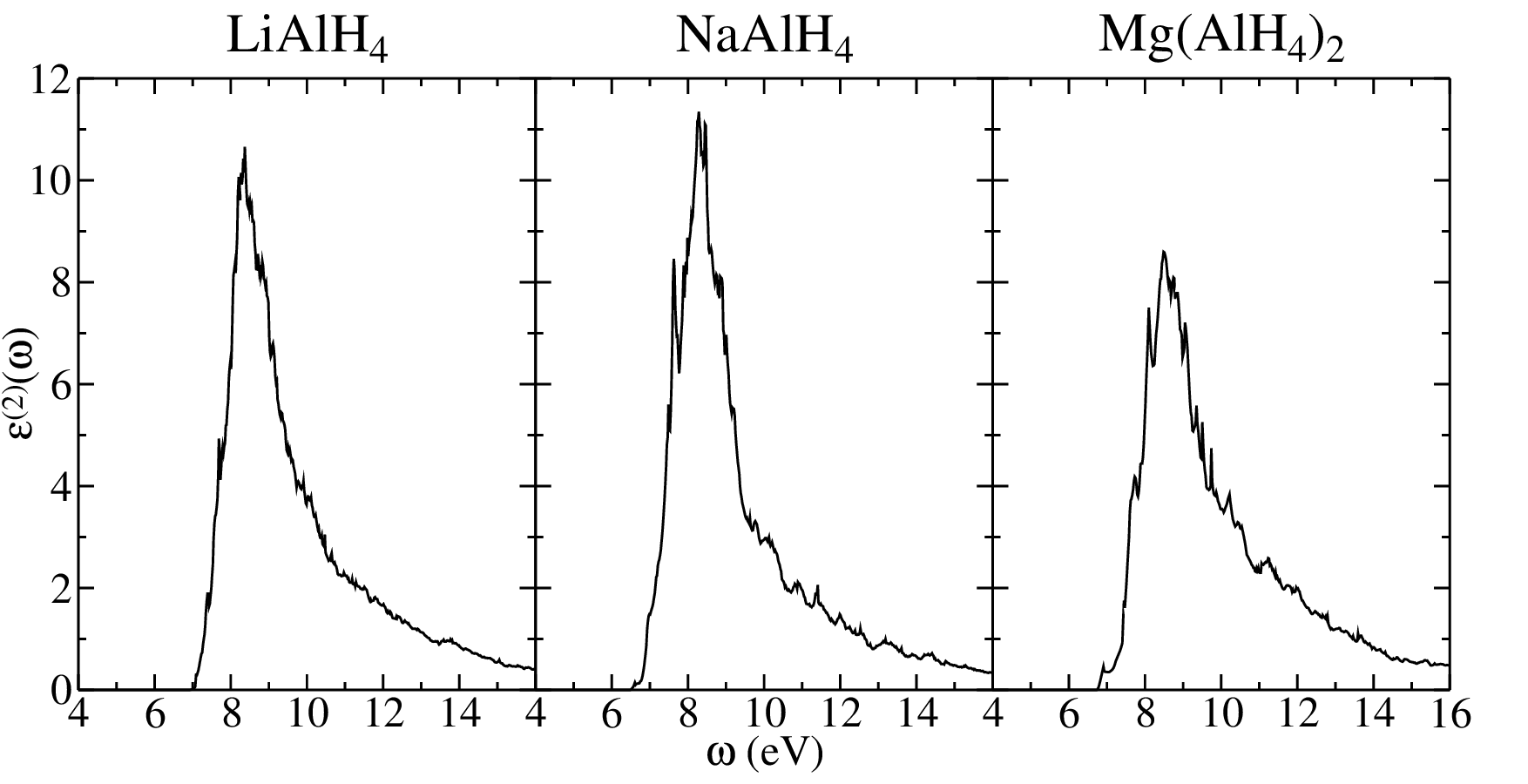}
\caption{\label{dieflinamg}Imaginary parts of the directionally averaged macroscopic dielectric
functions of $\lit$, $\naa$ and $\mg$.}
\end{figure}

In order to evaluate these differences we have also calculated the LDOS for the $\lit$ structure
with the Li$^+$~ions replaced by a uniform positive background. The LDOS on the Al atoms then
becomes very similar to that in the $\na$ and $\mg$ compounds. All these features indicate that
$\na$ and $\mg$ can be considered as ionic compounds, i.e., as a packing of AlH$_4^-$ anions and
Na$^+$ or Mg$^{2+}$ cations, whereas in $\lit$ there may be a stronger covalent contribution.

The calculated band gaps are given in Table~\ref{bingaps}. The $GW$ single particle band gaps of
the three compounds are almost the same and also in the optical gaps there is very little
difference.\cite{na_remark} This similarity indicates that the electronic structure around the band
gap is foremost determined by the $\alcl$ tetrahedra. The distances between these tetrahedra are
similar in these three compounds and apparently the detailed differences in their packing are
relatively unimportant.

This conclusion is strengthened by the dielectric function, which is shown in
Fig.~\ref{dieflinamg}. The maximal dielectric response of $\mg$ is somewhat smaller than that of
$\lit$ and $\na$, but the shape of the three curves is remarkably similar. The double peak
structure of the valence band of the LDOS, which appears in all three compounds in
Fig.~\ref{doslinamg}, is almost washed out in the dielectric response. Transitions from the lowest
valence band can be recognized only as a faint shoulder near 10~eV. It is then not surprising that
the smaller differences between the LDOS of $\lit$ and the other two compounds do not influence the
dielectric functions much.

\section{discussion}\label{discussion}
The electronic structure and the dielectric function of the binary compounds discussed in
Secs.~\ref{octalabel} and \ref{tetralabel} are foremost determined by the lattice of (AlH$_6)^-$
octahedra and $\alcl$ tetrahedra, respectively, whereas the cations have a minor influence. In this
section we will discuss this proposition in more detail. The exact value of the band gap is not
important in this discussion, only its relative variation with structure and composition. Since the
latter is described qualitatively by DFT/GGA calculations, see Table~\ref{bingaps}, we will only
use GGA results in this section.

\begin{figure}[!tbp]
\includegraphics[angle=270, width=8.6cm]{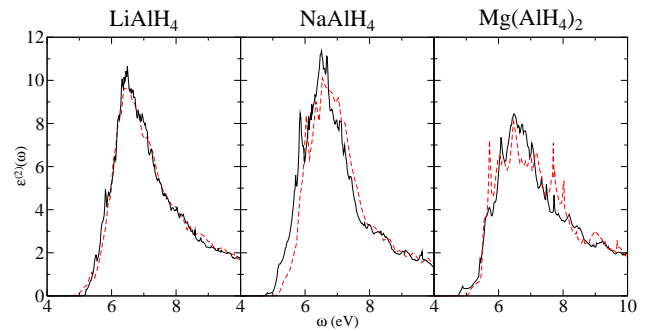}
\caption{\label{gem}(Color online) Dielectric functions of $\m$ compounds (black, solid) and of the
corresponding systems with Li$^+$, Na$^+$, and Mg$^2+$ ions substituted by a uniform background
charge (red, dashed). These results are based upon GGA calculations without scissors operator
cerrection.}
\end{figure}

We have calculated the dielectric functions of lattices of $\alcl$ tetrahedra in the $\lit$, $\na$
and $\mg$ structures, but with the Li$^+$, Na$^+$ and Mg$^{2+}$ cations replaced by a uniform
positive background charge. The results are compared to the dielectric functions of the real
compounds in Fig.~\ref{gem}. It can be observed that removing the cations in $\lit$ hardly changes
the dielectric function. Removing the cations in $\na$ results in a slight shift of the dielectric
response to higher energies. This is related to the disappearance of the peak at the bottom of the
conduction band, which has a sodium $s$ character, see the middle panel of Fig.~\ref{doslinamg}.
Also in $\mg$ removing the cations results in small changes in the dielectric function only. These
are mainly caused by the disappearance of the magnesium related peaks at the bottom of the
conduction band and a resulting flattening of the conduction bands, see the right panel of
Fig.~\ref{doslinamg}.

In conclusion, although removing the cations results in small changes in the conduction band,
overall the dielectric function changes very little, which means that it is foremost determined by
the lattice of $\alcl$ anions. Such a behavior is not uncommon for ionic compounds. For instance,
in alkali halides such as NaCl both the top of the valence band and the bottom of the conduction
band are determined by the anion lattice.\cite{boer1,boer2}

\begin{figure}
\includegraphics[width=8.3cm]{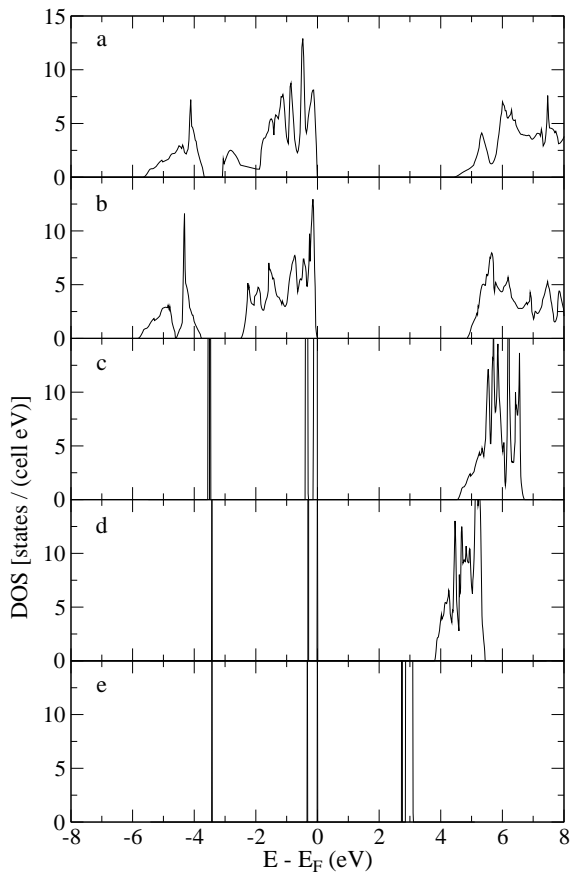}
\caption{\label{dossenmg}GGA densities of state of (a) $\mg$ and (b) with the Mg$^{2+}$ ions
replaced by a uniform background charge. While fixing the geometry of the $\alcl$ ions, the lattice
is expanded by a factor of 1,5 (c), 2 (d), and 5 (e).}
\end{figure}

Going one step further one can correlate the relative size of the band gap with the distance
between the anions. We will illustrate this using the DOS of $\mg$, which is shown in
Fig.~\ref{dossenmg}(a). Replacing the Mg$^{2+}$ ions by a uniform positive background does not
change the DOS significantly, as can be observed in Fig.~\ref{dossenmg}(b). In
Fig.~\ref{dossenmg}(c)-(e) the cell parameters are increased while the geometry of the $\alcl$
tetrahedra is fixed. As the distance between the anions increases, the band widths of all bands
decreases, but those of the valence bands decrease much more rapidly. At a large distance the DOS
is essentially that of an isolated $\alcl$ tetrahedron, where the valence states are split into
states of $s$ and $p$ symmetry due to the tetrahedral ligand field. The upper states of $p$
symmetry show a small splitting, since the tetrahedron has a small trigonal distortion. The gap
between the highest valence state and the lowest conduction state in the isolated tetrahedron is
only $\sim 0.5$ times the band gap in the solid, compare Figs.~\ref{dossenmg}(b) and
\ref{dossenmg}(e).

In general, the larger the distance between the anions, the smaller the gap. This result seems to
be somewhat counterintuitive, as at the same time the valence and conduction band widths are
decreasing and in general this would increase the gap. The result can be understood in terms of the
electrostatic (Madelung) potential.\cite{slater} Decreasing the lattice constant, the electrostatic
potential on the anions becomes more attractive to electrons, due to the closer packing of the
cations. The same argument also holds for a uniform positive background, i.e., a decreasing lattice
constant leads to a more attractive Madelung potential on the anions. All states on the anions
experience this potential and lower their energy. The size of this shift, however, depends upon the
degree of localization of the state. If a states is completely localized, the Madelung potential
will maximally lower its energy. On the other hand, if a state is completely delocalized, its
energy shift is zero, since the system as a whole is charge neutral.

The key point is that in our systems the valence states are much more localized than the conduction
states. This is immediately evident from Fig.~\ref{dossenmg}(b)-(e), where the valence band widths
decrease much faster with an increasing lattice constant than the conduction band widths. As a
result, the more localized valence states increase their energy significantly faster with an
increasing lattice constant than the conduction band states. Since this effect is much larger than
the effect of the decreasing band widths, increasing the lattice constant results in a smaller band
gap.

\begin{figure}
\includegraphics[width=5.0cm,angle=270]{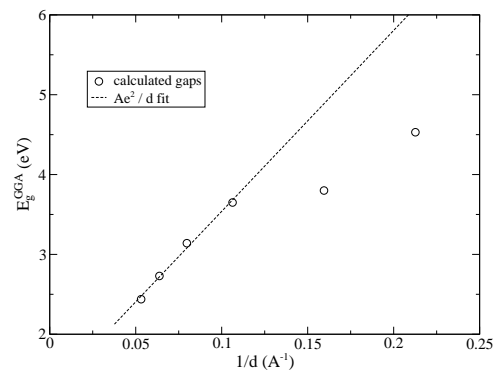}
\caption{\label{madelung}Calculated GGA band gaps of the expanded (AlH$_4$)$^-$-lattice (circles)
fitted to $Ae^2/d$ (dashed line), whith $d$ the distance between the anions; $A=1.57$.}
\end{figure}

This is quantified in Fig.~\ref{madelung}, where the band gaps of the 2, 3, 4, 5 and 6 times expanded
 lattice are fitted to an expression $Ae^2/d$, with $d$ the distance between the anions. The constant $A=1.57$
represents the difference in localization of the valence and conduction states. This simple model
breaks down if the localization of the states strongly depends upon $d$, i.e. in the 1.5 and 2 times expanded lattice. It occurs if $d$ becomes sufficiently small, see Fig.~\ref{dossenmg}(b).

This concept can be used to interpret the trend in the band gaps of the binary compounds, see
Table~\ref{bingaps}. In $\lit$, $\na$ and $\mg$ the Al-Al distance, which is a measure for the
distance between the anions, is almost the same, so their band gaps are very close. The Al-Al
distance in $\litt$ is much smaller than in $\naa$, which explains the larger band gap in the
former compound.


\section{Summary}

In this paper the electronic structures and dielectric functions of the simple hydrides $\lih$,
$\nah$, $\mgh$ and $\alh$, and the complex hydrides $\litt$, $\naa$, $\lit$, $\na$ and $\mg$, have
been studied by first principles calculations. The equilibrium structures of these compounds are
obtained from DFT/GGA total energy minimizations. $GW$ calculations within the QP approximation
provide the single particle excitation energies, i.e., the electronic band structures. We use the
$G_0W_0$ approximation based upon LDA wave functions and eigenvalues. The difference between the
dispersions of the $GW$ and the GGA bands is less than 10\%. Therefore, the band structures are
well represented by GGA band structures that are corrected by applying a scissors operator between
occupied and unoccupied states in order to obtain the $GW$ band gap. From the single particle wave
functions we then calculate the directionally averaged dielectric functions within RPA, neglecting
exciton effects. We also neglect local field effects, but from calculations on static dielectric
constants we conclude that this is a reasonable approximation.

All compounds are large gap insulators with band gaps that vary from 3.5 eV in AlH$_3$ to 6.5 eV in
the MAlH$_4$ compounds. In all cases the valence bands are dominated by the hydrogen atoms, whereas
the conduction bands have mixed contributions from hydrogen and metal cation states. The band gap
in LiH, AlH$_3$ and $\naa$ is direct, whereas in all the other compounds it is indirect. The
optical response of most compounds is qualitatively similar, notwithstanding sizeable differences
in their band structure and band gap. The dielectric function $\varepsilon^{(2)}(\omega)$ rises
sharply at photon energies corresponding to $\sim 6$ eV, and around $\sim 8$ eV it has a strong
peak reaching values in the range 10-15. In the direct gap materials $\varepsilon^{(2)}(\omega)$
has a weak tail going to lower energies. Between $\sim 8$ and $\sim 12$ eV,
$\varepsilon^{(2)}(\omega)$ gradually decreases to a value $\leq 2$ at 12 eV. Most of the materials
specific optical information can be found in this energy range, albeit in the form of relatively
weak shoulders in $\varepsilon^{(2)}(\omega)$.

The electronic structure and the optical properties of the aluminium compounds can be interpreted
in terms of aluminium hydride complexes, i.e., AlH$_6$ octahedra in AlH$_3$, $\litt$ and $\naa$,
and AlH$_4$ tetrahedra in $\lit$, $\na$. Explicit calculations on lattices of these complexes,
without the Li, Na, and Mg cations, show that the latter have a relatively small effect on the DOS
and on the optical response. The distance between the $\alcl$ tetrahedra in $\lit$, $\na$ and $\mg$
is almost the same. Since the interaction between the tetrahedra is then similar, this explains why
the optical spectra of these compounds are very similar.

The same reasoning can be applied to $\litt$ and $\naa$ in terms of a lattice of $\alcll$
octahedra. However, the distance between the octahedra is smaller in the Li compound because of the
smaller size of the cation. The band gap then becomes larger, which can be understood from the influence
of the increased Madelung potential on the more localized valence states.


\begin{acknowledgments}
The authors wish to thank R.~A.~de~Groot and P.~J. Kelly for helpful discussions, J.~Furthm\"uller
for the use of his optics package and G.~Kresse for making the linear response routines in VASP
available. This work is part of the research programs of `Advanced Chemical Technologies for
Sustainability (ACTS)' and the `Stichting voor Fundamenteel Onderzoek der Materie (FOM)'. The use
of supercomputer facilities was sponsored by the `Stichting Nationale Computerfaciliteiten (NCF)'.
These institutions are financially supported by `Nederlandse Organisatie voor Wetenschappelijk
Onderzoek (NWO)'.
\end{acknowledgments}

\bibliography{dfthydrogen,optixrem}

\begin{thebibliography}{77}
\bibitem[{QPr()}]{QPremark}
\bibinfo{note}{The QP energies $\epsilon_{n\mathbf{k}}$ are complex. The real
  parts comprise the band structure and are used as single particle excitation
  energies. The imaginary parts give the inverse QP lifetimes.}
\bibitem[{AR()}]{AR}
\bibinfo{note}{We use atomic radii $0.79$, $1.23$, $1.59$, $1.36$ and
  $0.81$~\Am for H, Li, Na, Mg and Al, respectively.}
\bibitem[{na3()}]{na3_remark}
\bibinfo{note}{In Ref.~\onlinecite{chou} a value of 4.6~eV is given for the
  single particle gap of $\naa$. This value is obtained by applying a $G_0W_0$
  shift to the GGA band gap, whereas we apply the shift to the LDA gap. The
  difference with our value of $3.94$ eV then mainly reflects the difference
  between the LDA and GGA gaps, see Table~\ref{bingaps}. Experience indicates
  that applying a $G_0W_0$ shift to the LDA gap generally results in a value
  that is within 10\% of the experimental single particle gap, see
  Refs.~\onlinecite{aryasetiawan,rubio}.}
\bibitem[{na_()}]{na_remark}
\bibinfo{note}{In Ref.~\onlinecite{chou} a value of 6.9~eV is given for the
  single particle gap of $\na$. Again this value is obtained by applying a
  $G_0W_0$ shift to the GGA band gap.}
\end{thebibliography}

\end{document}